\documentclass[prc,aps,showpacs]{revtex4}
\usepackage{graphicx}
\usepackage{subfig}
\usepackage{float}
\begin{document}

\title{Spin-orbit decomposition of \textit{ab initio} nuclear wavefunctions}

\author{Calvin W. Johnson}
\affiliation{Department of Physics, San Diego State University,
5500 Campanile Drive, San Diego, CA 92182-1233}
\affiliation{Computational Sciences Research Center, San Diego State University,
5500 Campanile Drive, San Diego, CA 92182-1245}

\pacs{21.60.Cs,21.60.De,27.20.+n,21.10.Re}

\begin{abstract}
Although the modern shell-model picture of atomic nuclei is built from single-particle orbits 
with good total angular momentum $j$, leading  to $j$-$j$ coupling, decades ago phenomenological 
models suggested a simpler picture  for $0p$-shell nuclides can be 
realized via coupling of total spin $S$ and 
total orbital angular momentum $L$. I revisit this idea with large-basis, 
no-core shell model (NCSM) calculations using modern \textit{ab initio} two-body interactions, and dissect 
the resulting wavefunctions into their component $L$- and $S$-components.  Remarkably, there is 
broad agreement with calculations using the phenomenological Cohen-Kurath forces, despite a gap of 
nearly fifty years and six orders of magnitude in basis dimensions.  I suggest $L$-$S$ 
decomposition may 
be a useful tool for analyzing \textit{ab initio} wavefunctions of light nuclei, for example 
in the case of rotational bands.
\end{abstract}

\maketitle

\section{Introduction}

Microscopic theories of low-energy nuclear structure arguably began with the realization that one could base nuclear wavefunctions on an \textit{independent particle model} (IPM),  with the crucial proviso
that one has a strong 
spin-orbit force. The IPM was motivated by filled shells (`magic numbers') and by the magnetic moments of nuclei with one particle 
outside or one hole in a closed shell \cite{May48,HJS49,FHN49}.  Even today many high-end methods for \textit{ab initio} nuclear structure, such as Green's Function 
Monte Carlo \cite{CarlsonGFMC,GFMC}, coupled-clusters \cite{DHJ04,HPDHJ10}, and the no-core shell model \cite{NVB00}, use
the IPM as a starting point, although each go far beyond it.

Because the nuclear Hamiltonian is rotationally invariant, total angular momentum $J$ is a good quantum number 
(as is the third or $z$ component, $M$). In the nuclear IPM, despite the lack of a core as in atomic physics, one uses an 
average or mean potential, typically one which  is rotationally invariant, 
 to construct the single-particle states.   These single-particle states have good orbital angular momentum $l$ and 
intrinsic spin $s$, which for electrons and for nucleons is $(1/2)\hbar$, and following the 
rules for addition of quantized angular momentum \cite{Edmonds}, symbolized by $\oplus$, one can combine these into the total 
angular momentum for a single particle, 
\begin{equation}
j= l \oplus s  \label{ls1p}.
\end{equation}
Single particle states which are degenerate or nearly so are grouped together into shells, and the IPM is often called the 
\textit{non-interacting shell model}. 

For atoms, with weak coupling between orbital angular 
momentum and spin, single-particle states with the same $l$ but different $j$ are nearly degenerate. In that case it 
makes sense, to follow $L$-$S$ or Russell-Saunders coupling and  couple together all the individual orbital angular momenta 
for $A$ particles, 
\begin{equation}
L = l_1 \oplus l_2 \oplus l_3 \oplus \ldots l_A,
\end{equation}
into total orbital angular momentum $L$, and similarly for the total spin $S$, and then construct
\begin{equation}
J = L \oplus S.  \label{LSmanyp}
\end{equation}
Indeed, such $L$-$S$ coupling could be found in the first approaches to the IPM for nuclei \cite{FW37,FP37}.   With the 
understanding of a strong nuclear spin-orbit coupling, however, it was seen as advantageous to adopt 
$j$-$j$ coupling \cite{Flo52,Kur52}, by first coupling up $l$ and $s$ for each particle as in (\ref{ls1p}) and then summing the 
$j$s:
\begin{equation}
J = j_1 \oplus j_2 \oplus j_3 \oplus \ldots \oplus j_A.
\label{jjmanyp}
\end{equation}

Nowadays the IPM has been superseded by 
the \textit{interacting}  or  configuration-interaction 
(CI) shell model \cite{BG77,BW88,ca05} and other many-body methods.  Nonetheless one can consider how good the IPM is as a starting point by 
 looking at semi-magic shells. For example, using the phenomenological interaction KB3G (which is a 
monopole-adjusted version\cite{Pov01} of the Kuo-Brown interaction\cite{KB68} ) in the $pf$-shell, one finds the full configuration-interaction 
wavefunction of $^{48}$Ca is $90\%$ a filled $(0f_{7/2})^8$ configuration; and in the $sd$-shell, using the phenomenological 
USDB (universal $sd$-shell interaction, version B) interaction\cite{USDB}, the ground state of 
$^{24}$O is $91\%$ filled $(0d_{5/2})^6 (1s_{1/2})^2$ 
configuration, and the ground state of $^{22}$O is $75\%$ filled $(0d_{5/2})^6$ 
configuration.  

This simple success is not universal. In the $sd$-shell, with the same USDB phenomenological 
interaction, $^{28}$Si is only $21\%$ a filled  $(0d_{5/2})^{12}$ configuration,
 and in the $p$-shell, using the Cohen-Kurath interaction \cite{CK65}, the ground state of 
$^8$He is only $37\%$ a filled $(0p_{3/2})^4$ configuration while $^{12}$C is about $51\%$ a filled $(0p_{3/2})^8$ configuration.
In fact, it was known long ago, at least phenomenologically \cite{Ing53,Kur56}, 
that $p$-shell nuclei are intermediate between  $j$-$j$ and $L$-$S$  coupling.  In some $p$-shell cases the latter leads to a 
simpler description: the ground state of  $^8$He is $96\%$ $L=0$ and $^{12}$C is $82\%$ $L=0$
(see also section 5 of \cite{Mil09}).  For heavier nuclei, 
with stronger spin-orbit forces, $L$-$S$ coupling is less satisfactory: 
for the $sd$-shell cases given above have grounds states which are roughly only $35 \% L=0$
components, and the $^{48}$Ca ground state has only about $20\%$ $L=0$ component. (How
these decompositions are carried out will be  described in more detail below.)

This suggests $L$-$S$ decomposition as a tool to investigate theoretical wavefunctions, even if it is not directly experimentally 
measureable,  especially  for $p$-shell nuclei where phenomenological interactions suggest strong dominance by a few 
$L$-$S$ components.  But what about more ``realistic'' interactions? The past two decades have seen tremendous advances in 
\textit{ab initio} calculations of nuclear structure, mostly for $p$-shell nuclides.  

As I will show below, both the phenomenological and \textit{ab initio} wavefunctions, despite separated by six orders of magnitude 
in the basis dimensions, and over four decades in the origin of the interactions, show remarkable congruence in their 
$L$-$S$ decomposition.  While $L$-$S$ composition is not directly measureable, it does have an effect on transitions such as 
Gamow-Teller, M1, and so on; furthermore this congruence suggests that both the old and the new calculations are probably doing 
something right.

Of course, any spin-orbit force will play a big role in the $L$-$S$ decomposition, and it is well-known that \textit{ab initio} three-body 
forces strongly influence spin-orbit interactions, for example to get the correct ground state spin for some $p$-shell nuclides 
\cite{NO03}.  In this  paper, intended as an introduction to and demonstration of 
the method, I only look at two-body interactions, but in the near future will
examine how these results change with the addition of three-body forces. 

\section{Methods}
\label{methods}

The tools used for this investigation come in three parts: the many-body method; the interactions used; and decomposition 
of the wavefunctions into $L$ and $S$ components.

The many-body method I use is configuration-interaction (CI) diagonalization of the many-body Hamiltonian in a 
shell-model basis \cite{BG77,BW88,ca05}, using the BIGSTICK code \cite{BIGSTICK}. 
 Here one defines a finite 
single-particle space and has as input single-particle energies and two-body matrix elements; three-body interactions can 
also be used, but will be investigated in future work.  The calculations are carried out in occupation space, with 
occupation-representation of Slater determinants built from single-particle states as the many-body basis states.  
In brief, one chooses a finite set of single particle orbits with good orbital angular momentum $l$ (and thus good parity) and 
good angular momentum $j$ and $z$-component $m$; with that it easy to build basis states with fixed total $M$; thus 
BIGSTICK is termed an M-scheme code.  BIGSTICK computes the many-body Hamiltonian in this basis from the input 
interaction matrix elements, i.e., it computes 
\begin{equation}
H_{ab} = \langle \Psi_a | \hat{H} | \Psi_b \rangle,
\end{equation}
where $ | \Psi_a \rangle $ is an occupation representation of a Slater determinant with fixed $M$,
and finds the low-lying eigenpairs 
\begin{equation}
\mathbf{H} \vec{v}_\lambda = E_\lambda \vec{v}_\lambda
\end{equation}
by the Lanczos algorithm \cite{GVL,Lanczos}. 

Within this framework I use two different model spaces and interactions.  First is the phenomenological Cohen-Kurath 
interaction \cite{CK65}, which works entirely within the $0p_{3/2}$-$0p_{1/2}$ space; it has two single-particle energies and 15 
unique two-body matrix elements. Because the matrix elements were fitted to spectra, the radial component of the single-particle wavefunctions have not been rigorously defined; fortunately for my purpose they are not needed.  For a given number of 
valence protons and neutrons, and for fixed total $M$, all possible configurations are used.

The second are \textit{ab initio} interactions in the so-called no-core shell model (NCSM) framework \cite{NVB00}. Here one uses 
harmonic oscillator single-particle states with a fixed frequency $\Omega$, and 
utilizes the $N_\mathrm{max}$ truncation on the many-body states: one allows 
only many-body harmonic oscillator configurations which are a maximum of $N_\mathrm{max}\hbar\Omega$ in energy above the 
lowest harmonic oscillator configuration.  This allows one to exactly decouple the relative wavefunction from center-of-mass 
motion \cite{Pal67,PP68,GL74}, although that is not  important to this study.

High-precision \textit{ab initio} interactions are fitted to low-energy nucleon-nucleon scattering phase shifts and to deuteron properties.
Among the first was the Argonne V18 \cite{Argonne}, while more recent ones, such as the one used in this study \cite{EM03},
are derived from chiral effective field theory \cite{Wei90,Wei91,OvK92,BvK02}.

These interactions generally have a large coupling between high- and low-momentum components, which is often interpreted 
as a `hard core;' such a hard core can be seen directly in local interactions fitted to scattering data such as the Argonne V18 
and related potentials. While such interactions can be used directly in coupled-cluster  (CC) calculations \cite{DHJ04,HPDHJ10} and, when local or 
nearly so, in Green's function Monte Carlo  (GFMC) calculations \cite{CarlsonGFMC,GFMC}, both of those very powerful methodologies 
favor the ground state. Finding 
excited states are challenging, though not impossible, for CC and GFMC calculations, and the technology for projecting out 
the $L$ and $S$ components in those methods has not yet been developed. 

For configuration-interaction (CI) calculations, on the other hand, 
obtaining excited states and decomposition into $L$ and $S$ components 
is hardly more difficult than finding the ground state.   On the other hand, unlike CC calculations, CI calculations include unlinked diagrams \cite{Sh98}, and because of the strong coupling 
between low- and high-momentum states, the basis for CI grows exponentially 
and convergence with the size of the space is very slow in all but the smallest nuclides. Therefore for  \textit{ab initio} CI calculations in computationally 
tractable spaces one usually
softens or renormalizes the interaction via a unitary transformation.  A very popular unitary transformation is the 
similarity renormalization group (SRG) \cite{GW93,Weg94,BFP07, BFS10}, whereby the Hamiltonian is evolved by a flow equation:
\begin{equation}
\frac{d\hat{H}(s)}{ds} = \left [ \hat{\eta}(s), \hat{H}(s) \right ],
\end{equation}
where one commonly chooses the generator of the flow to be $\hat{\eta}(s) = \left [\hat{T}, \hat{H}(s) \right ],$
with $\hat{T}$ the kinetic energy; this drives the Hamiltonian in momentum space towards the diagonal and weakens the coupling 
between low- and high-momentum states. 
 Because the transformation is unitary, any quantity represented by an eigenvalue, 
such as scattering phase shifts or the on-shell T-matrix elements if evolved in free space, remains unchanged. What do change 
are, for example, the off-shell T-matrix elements, but exploring that topic further is beyond the focus of this paper; furthermore, 
3-body forces are induced by the evolution \cite{JNF09,JMF13}. 
I follow the convention of parameterizing the evolution not by $s$ but by $\lambda =( m_N^2 / \hbar^4 s)^{1/4}$, with 
$m_N$ the nucleon mass; then $\lambda$ has units of fm$^{-1}$.  Other authors follow $s$, which is sometimes written as 
$\alpha$.


\subsection{$L$-$S$ decomposition}
\label{ls}

It is worth describing in a little detail how the $L$-$S$ decomposition is carried out.  Suppose one wants to 
find the fraction of a CI wavefunction with a given $L$, that is, to expand 
\begin{equation}
| \Psi \rangle = \sum_L c(L) | L \rangle \label{L_decomp}
\end{equation}
but in general, the dimension of the subspace of states with a given $L$ is greater than one. One naive method then is 
to generate  all many-body states of a given $L$ or $S$, that is, $\{ | L; a \rangle \}$ where $a$ carries any additional 
information needed to label such states. Then the fraction of a state with that $L$ is
\begin{equation}
|\langle L | \Psi \rangle|^2 = \sum_{a \in L} | \langle L; a | \Psi \rangle |^2.
\end{equation}
This presupposes one can generate all the  $\{|L; a \rangle \}$, for example by 
diagonalize the operator $\hat{L}^2$, but that is easy only in small spaces.

Instead I turn to a modification of the `Lanczos trick', invented for generating strength functions \cite{CPZ90,CMN99,HNZ05} and which 
has been previously used to analyze phenomenological states in terms of their SU(3) irreps \cite{GDJ00}.

Let  $|\Psi \rangle$ be a CI wavefunction I wish to decompose in components labled by the eigenvalues of a Hermitian operator, 
in this case $\hat{L}^2$ .  I carry out the Lanczos algorithm with $|\Psi\rangle$ as my \textit{pivot}, that is the 
starting vector $|v_1 \rangle$: 
\begin{equation}
\begin{array}{cccccc}
\hat{L}^2 |v_1 \rangle & = & \alpha_1 | v_1 \rangle  &  + \beta_1 | v_2 \rangle & &   \\
\hat{L}^2 |v_2 \rangle &= & \beta_1 | v_1 \rangle & + \alpha_2 | v_2 \rangle & +  \beta_2 | v_3 \rangle &\\
\hat{L}^2 |v_3 \rangle &= &  & \beta_2 | v_2 \rangle & + \alpha_3 | v_3 \rangle & +  \beta_3 | v_4 \rangle \\
\ldots & &  & & & \nonumber
\end{array}
\end{equation}
As is well-known for the Lanczos algorithm \cite{Lanczos}, this procedure generates a Krylov subspace and the eigenvalues of 
the tridiagonal matrix given by $\alpha_i, \beta_i$ will converge to the extremal eigenpairs of $\hat{L}^2$, of which 
the eigenvectors are a linear combination of the Lanczos vectors,
\begin{equation}
| L \rangle = \sum_i d_i(L) | v_i \rangle,
\end{equation}
which is an inversion of (\ref{L_decomp}).
In fact the Krylov space is exhausted by the eigenvectors of $\hat{L}^2$ contained in the pivot, so that to get the fraction 
of $|\Psi \rangle=|v_1\rangle$ with orbital angular momentum $L$, 
it is just
\begin{equation}
 | \langle L| \Psi \rangle |^2=  | d_1(L)|^2
\end{equation}
so that no sum is needed and one can simply read off the amplitude. 

One can decompose using any Hermitian operator, and the procedure for decomposition 
with spin $S$ is identical to the above.    One can do joint
decomposition, that is decompose a wavefunction into states of specific $L$ and $S$, 
but I do not carry out such fine-grained analysis here.

\section{Results}
\label{results}

I give results for four nuclides from roughly the middle of the $p$-shell: $^9$Be, $^{10,11}$B, and $^{12}$C. 
In particular I look at states in $^{12}$C known to be problematic, and at rotational bands in $^9$Be. 
For all NCSM calculations I used $N_\mathrm{max} = 6$ (chosen so all the 
calculations could be easily carried out on a desktop computer using the BIGSTICK code), 
an oscillator frequency $\hbar\Omega = 20$ MeV for $^9$Be and 22 MeV for the other nuclides 
for the single-particle basis, which roughly minimized the ground state energies, and an SRG evolution 
parameter of $\lambda = 2.0$ fm$^{-1}$.  I also carried out an $N_\mathrm{max}=8$ calculation for $^{10}$B 
which required supercomputer time, for reasons discussed below in section \ref{b10}.
Because the 
phenomenological calculations with the Cohen-Kurath force only include ``normal'' parity states, i.e., the same 
parity as the ground state, I only show those, although 
there is no barrier to dissecting unnatural parity states using the method described in Section \ref{ls}

In standard NCSM procedures one carefully finds the variational minimum as a function of 
$\hbar \Omega$ and studies the convergence as a function of the model space parameter 
$N_\mathrm{max}$.   Instead, I demonstrate  selected $^{12}$C results are relatively 
robust under variation of both $\hbar \Omega$ and $\lambda$. Using a fifth light nuclide, 
$^7$Li, I demonstrate robustness as $N_\mathrm{max}$ is varied from 6 to 12.   
Thus one can have good confidence in the general results obtained here.

\begin{figure}
\centering
\includegraphics[scale=0.45,clip]{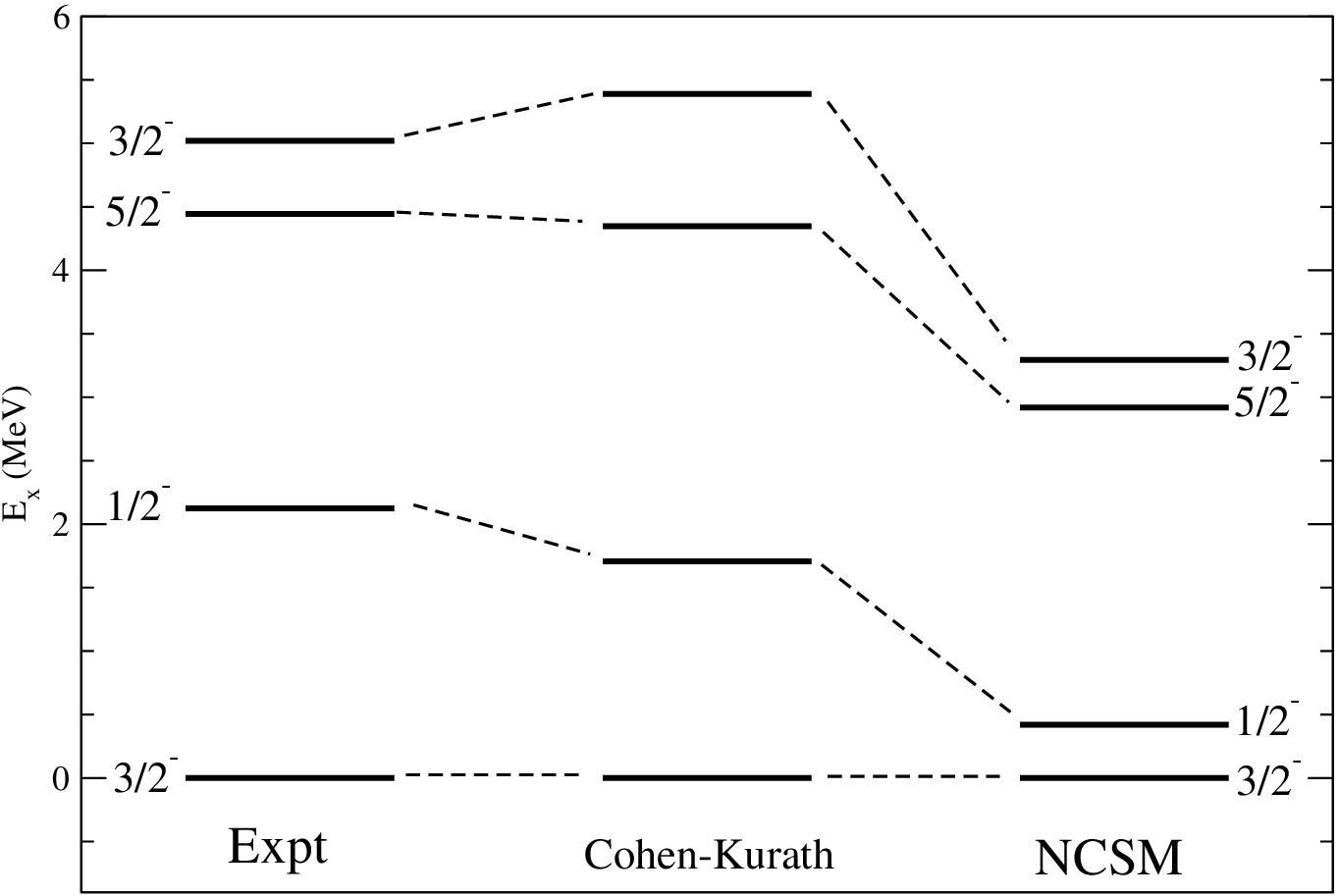}
\caption{ Low-lying excitation spectrum of $^{11}$B, comparing experiment, 
the Cohen-Kurath interaction, and the no-core shell model (NCSM) using an chiral two-body force 
evolved via SRG to $\lambda= 2.0$ fm$^{-1}$  .  All the states have $T=1/2$.}
\label{B11spect}
\end{figure}



\subsection{$^{11}$B}
Let's begin with an odd-$A$ nucleus.
 Fig.~\ref{B11spect} compares the low-lying excitation energies 
from experiment (experimental spectra for all cases in this paper are taken from the  National Nuclear 
Data Center \cite{nndc}) , from the Cohen-Kurath interaction in the $0p$ space, with a dimension of 62, and a NCSM calculation, 
with a dimension of 20 million, using $\lambda =2$ fm$^{-1}$ and $\hbar \Omega=22$ MeV; the latter choice approximately 
minimizes the ground state energy in this space. 
All the low-lying states have $T=1/2$, exactly in the case of Cohen-Kurath; the NCSM interaction 
includes isospin breaking terms but for all the cases in this paper the isospin assignments are very good.

\begin{figure}
\centering
\includegraphics[scale=0.45,clip]{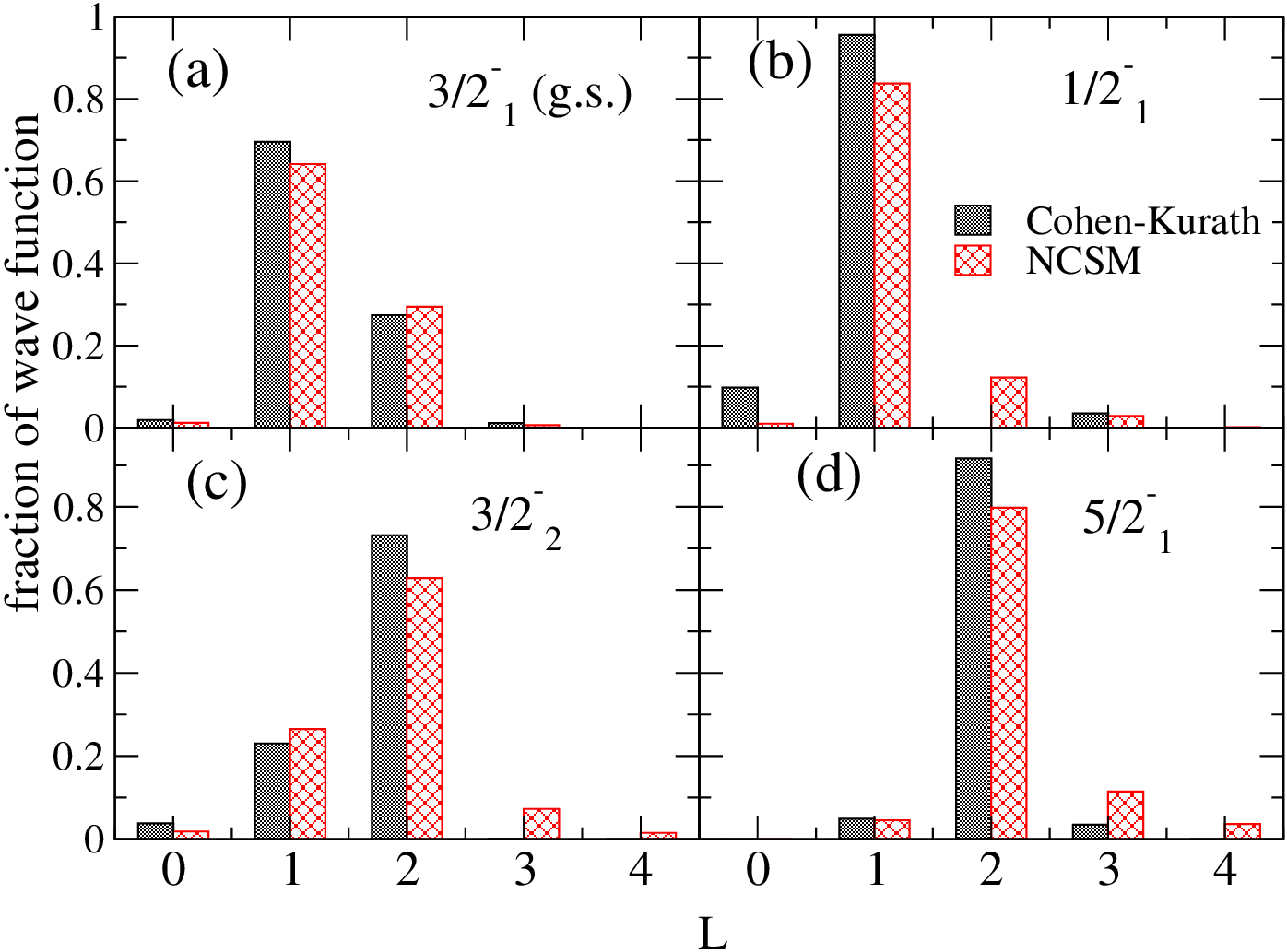}
\caption{(Color online) Decomposition of  low-lying states of $^{11}$B into 
components of good $L$ (total orbital angular momentum), comparing wavfunction 
computed from
the Cohen-Kurath interaction (black/dark shaded), and from the NCSM (red/cross-hatched).   
All the states have $T=1/2$.}
\label{B11_Ldecomp}
\end{figure}

\begin{figure}
\centering
\includegraphics[scale=0.45,clip]{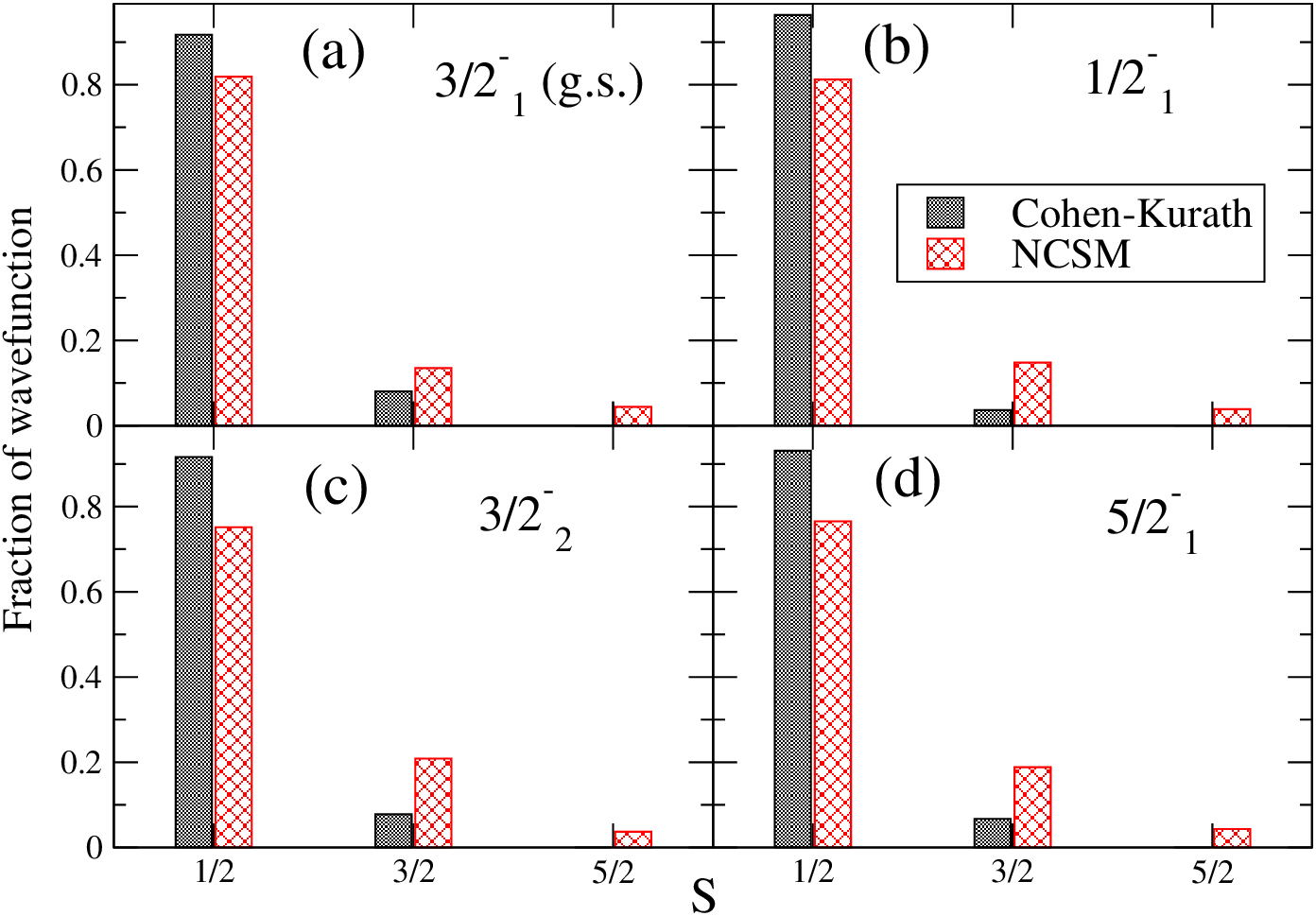}
\caption{(Color online) Decomposition of  low-lying states of $^{11}$B into 
components of good $S$ (total spin), comparing wavfunction 
computed from
the Cohen-Kurath interaction (black/dark shaded), and from the NCSM (red/cross-hatched).   
All the states have $T=1/2$.}
\label{B11_Sdecomp}
\end{figure}
The $L$-decomposition is displayed in Fig.~\ref{B11_Ldecomp}. 
 Here both the 
first and second $3/2^-$  show contrasting patterns, with 
agreement between Cohen-Kurath and the NCSM.  I also show the spin $S$-decomposition in 
Fig.~\ref{B11_Sdecomp}, which displays good qualitative agreement between the two calculations.

One can take this kind of decomposition farther, for example decompose the spin $S$ into the proton and 
neutron components $S_p$ and $S_n$, respectively. The low-lying states discussed here are all dominated ($ > 80\%$) by 
an $S_p = 1/2$, coupling primarily to $S_n =0$ to form $S=1/2$ and $S_n=1$ to form $S=3/2$. Both the 
NCSM calculation and the phenomenological Cohen-Kurath agree, with the exception of the $S=3/2$ component of
 the second $3/2^-$ state; here the Cohen-Kurath wavefunction is roughly equally divided between the two subcomponents, 
but I note that in this space  four neutrons form only two $S=0$ states and 3 $S=1$ states, severely constraining the results.

\subsection{$^{10}$B}
\label{b10}

\begin{figure}
\centering
\includegraphics[scale=0.45,clip]{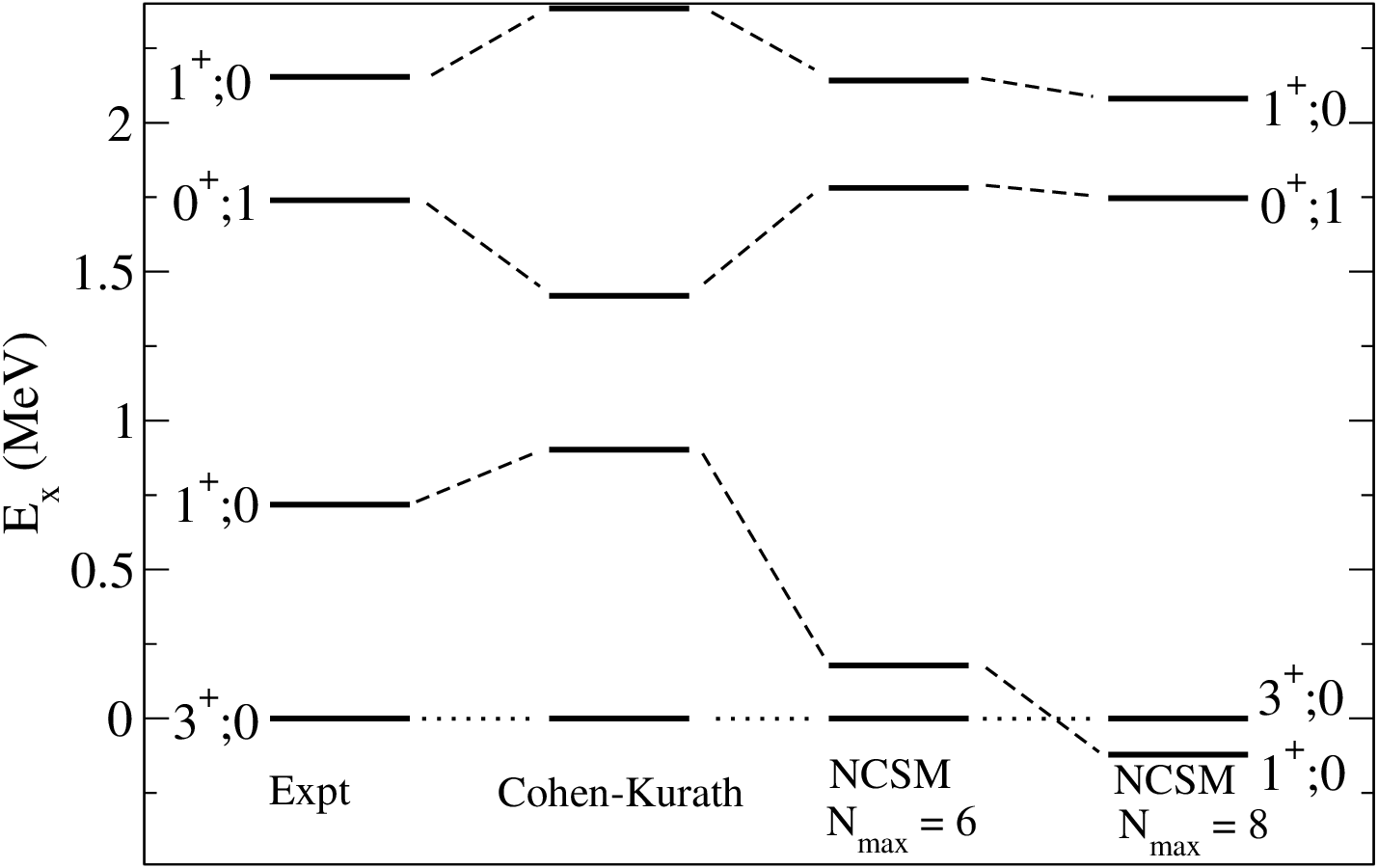}
\caption{Low-lying excitation spectrum of $^{10}$B, comparing experiment, 
the Cohen-Kurath interaction, and the no-core shell model (NCSM) using an chiral two-body force 
evolved via SRG to $\lambda= 2.0$ fm$^{-1}$,, with the harmonic oscillator basis frequency 
$\hbar \Omega =22$ MeV, for $N_\mathrm{max} =6,8$ .}
\label{B10spect}
\end{figure}
\begin{figure}
\centering
\includegraphics[scale=0.45,clip]{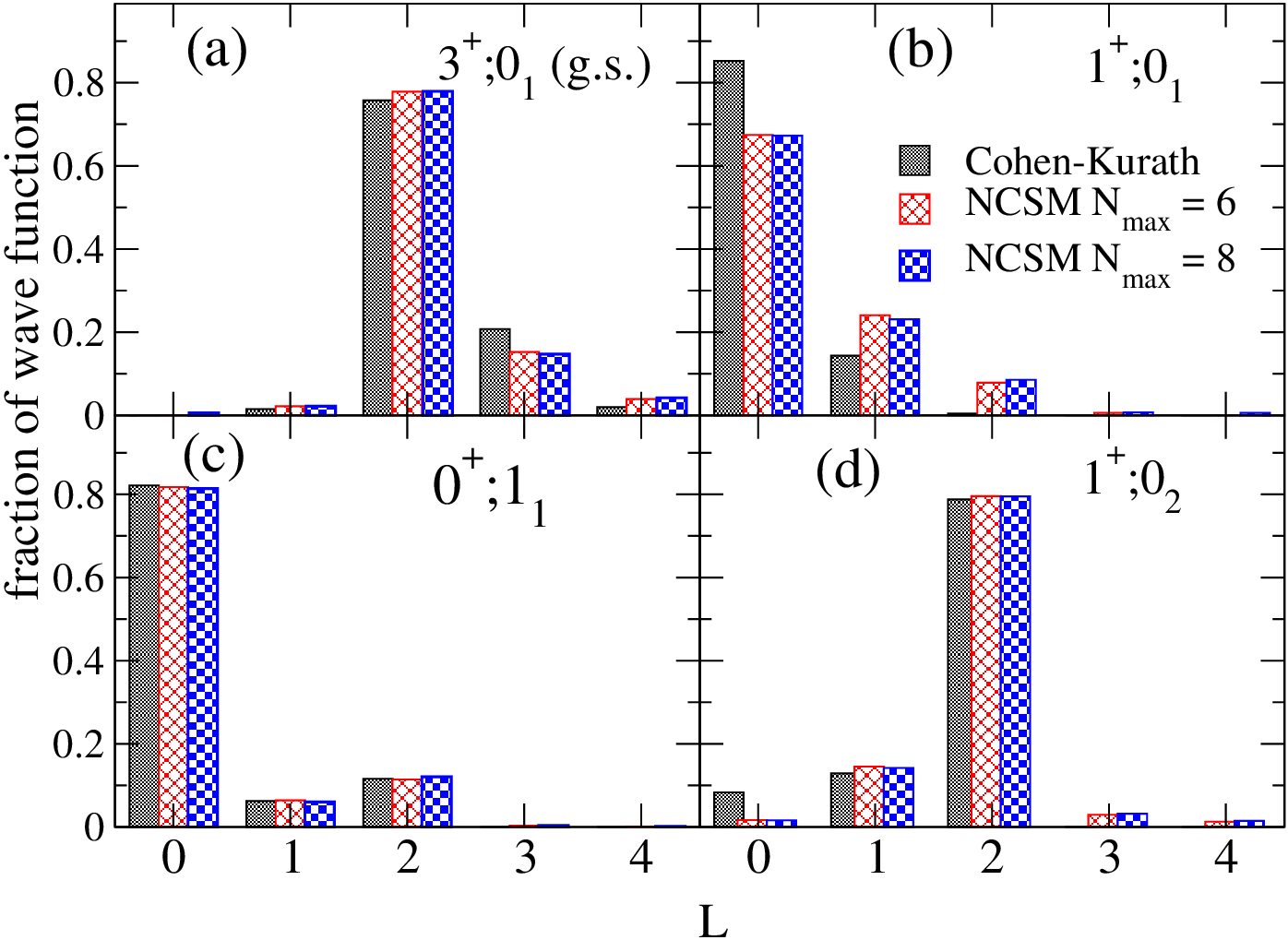}
\caption{(Color online) Decomposition of  low-lying states of $^{10}$B into 
components of good $L$ (total orbital angular momentum), comparing wavfunction 
computed from
the Cohen-Kurath interaction (black/dark shaded), and from the NCSM for both $N_\mathrm{max}=6$ (red/cross-hatched)
and  $N_\mathrm{max}=8$ (blue/checkered).}
\label{B10_Ldecomp}
\end{figure}

Now I turn to the case of $^{10}$B.  Prior work demonstrated that \textit{ab initio} calculations with two-body forces 
alone yielded the wrong ground state spin, and only the introduction of three-body forces produced the corrected ordering of the low-lying states 
\cite{NGV07,JMF13}.   For this work I carried out NCSM calculations (2-body only) both at $N_\mathrm{max} =6$ and 8, with dimensions of 12 million and 
166 million, respectively, using $\lambda =2$ fm$^{-1}$ and $\hbar \Omega=22$ MeV. These spectra, along with the experimental spectrum and 
from  the Cohen-Kurath interaction in the $0p$ space, with a dimension of 84, are shown in  Fig.~\ref{B10spect}.  Although the $3^+$ state is the ground 
state in the $N_\mathrm{max}=6$ calculation, the $1^+$, which is well known to be slow to converge \cite{JMF13}, 
drops below it for the $N_\mathrm{max}=8$ calculation.

Figure \ref{B10_Ldecomp} shows the decomposition of selected states into their 
components with good $L$. 
Of particular interest are the first and second $1^+;0$ states, which show contrasting patterns ($1^+;0_1$ is dominated by 
$L=0$ while $1^+;0_2$ is dominated by $L=2$), with both the  the Cohen-Kurath and NCSM wavefunctions 
giving good agreement, even in minor components, despite the vast difference in model space sizes and the origin of the forces. Note 
that even though the $1^+_1$ state drops below the $3^+$ as one goes from $N_\mathrm{max}=6$ to $N_\mathrm{max}=8$, the 
$L$-decomposition is nearly identical. 
  Other low-lying 
states show similar agreement.  The decomposition according to 
spin $S$ is of similar quality and not shown.

Agreement between the Cohen-Kurath and the \textit{ab initio} NCSM calculations does not mean they are both \textit{right}, 
but it does certainly bolster  confidence in the calculations.  Below, in the case of $^{12}$C, I will show some cases where 
there are discrepancies, which happen to occur in states known to be problematic.

\subsection{$^{12}$C}

\begin{figure}
\centering
\includegraphics[scale=0.45,clip]{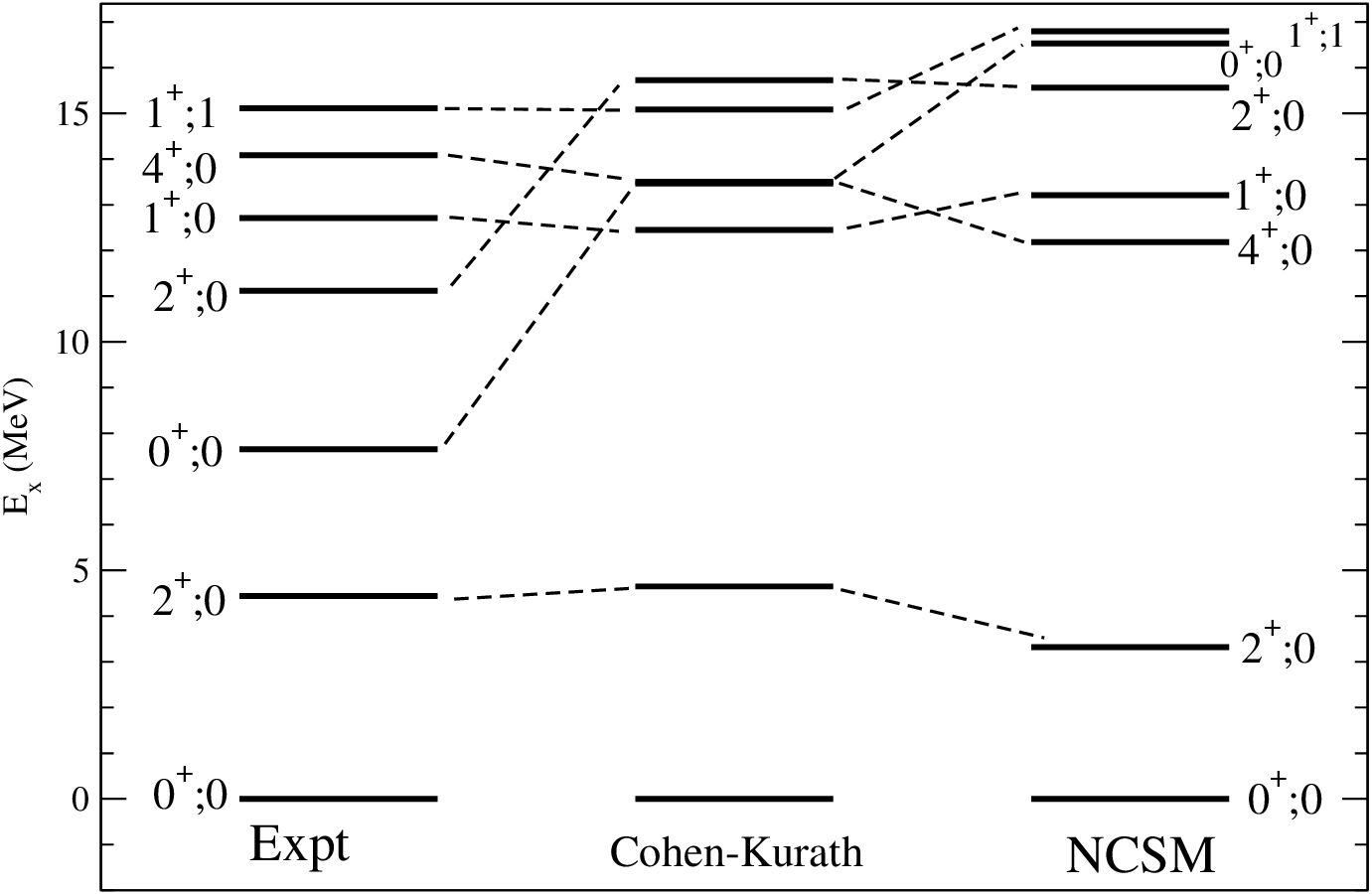}
\caption{ Low-lying excitation spectrum of $^{12}$C, comparing experiment, 
the Cohen-Kurath interaction, and the no-core shell model (NCSM) using an chiral two-body force 
evolved via SRG to $\lambda= 2.0$ fm$^{-1}$, with a harmonic oscillator basis frequency of 
$\hbar \Omega = 22$ MeV.}
\label{C12spect}
\end{figure}

Of particular interest is $^{12}$C, in part because it it so difficult to get its spectrum correct. As seen in Figure \ref{C12spect}, 
neither the phenomenological Cohen-Kurath calculation nor the NCSM calculations get the second $0^+$ state and the subsequent 
band near the correct energy. This is the famous Hoyle state \cite{Hoyle} and is known to have predominantly four-particle, four-hole structure, with similar 
states found in $^{16}$O.  Recent calculations have suggested that, in a harmonic oscillator basis, the Hoyle state mixes in many states of high $N$ \cite{Neff12}, 
making it difficult to access in standard CI shell-model calculations.  

The Hoyle state is far from the only problem,  
not least because recent calculations \cite{MVC14} get the excitation energy of the 
first $1^+;0$ state wrong while obtaining a good value of the $B(M1)$ from the first $1^+;1$ state to the ground state. 
This problem in particular inspired this work. 

\begin{figure}
\centering
\includegraphics[scale=0.45,clip]{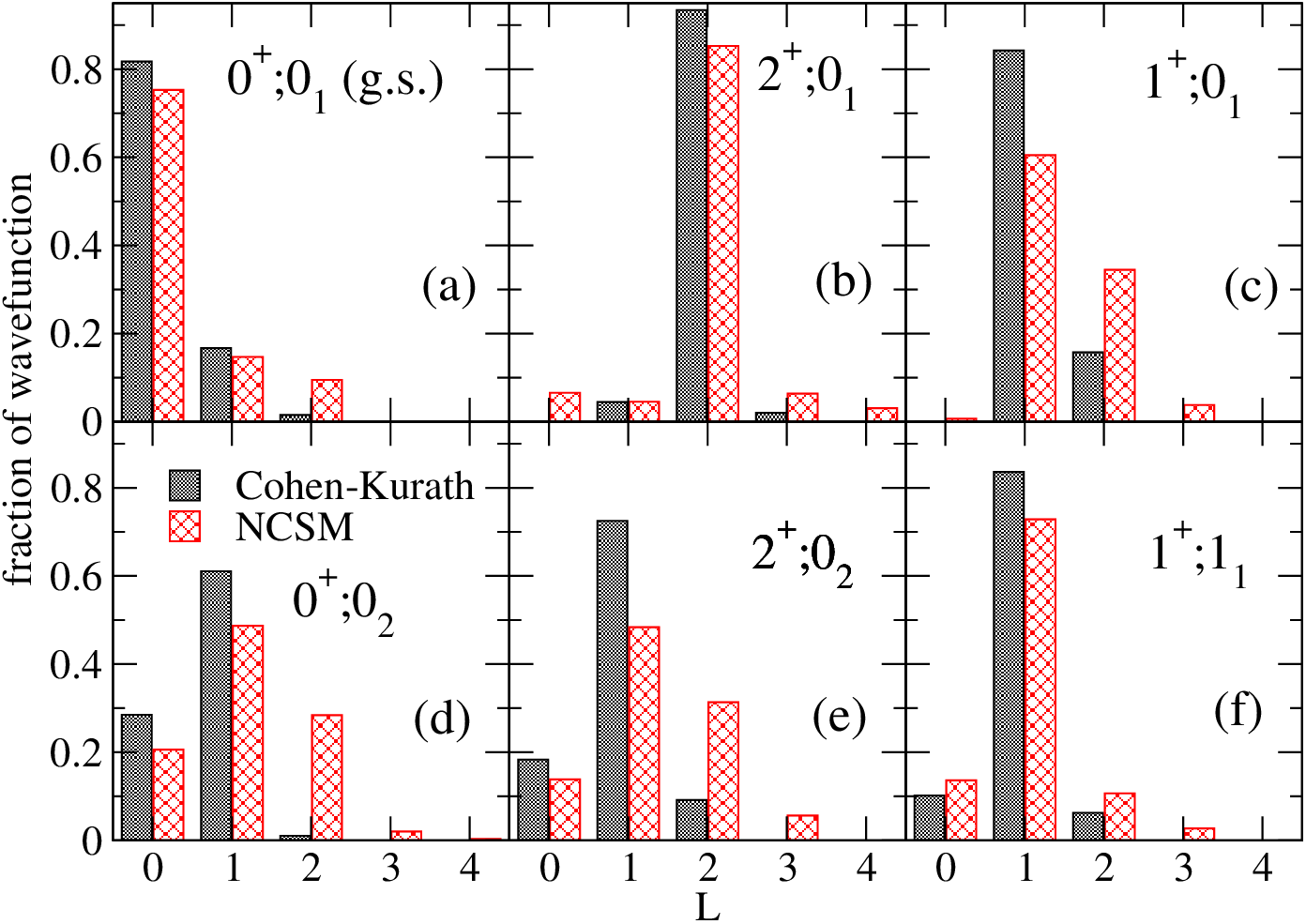}
\caption{(Color online) Decomposition of  low-lying states of $^{12}$C into 
components of good $L$ (total orbital angular momentum), comparing wavfunction 
computed from
the Cohen-Kurath interaction (black/dark shaded), and from the NCSM (red/cross-hatched).}
\label{C12_Ldecomp}
\end{figure}

Fig.~\ref{C12_Ldecomp} shows the $L$-decomposition of the ground state band $0^+$ and $2^+$ states, as well as 
the excited band $0^+_2$ and $2^+_2$ and the first $1^+;0$ and $1^+;1$ states.  While the ground state band finds 
agreement between Cohen-Kurath and NCSM, and is rather simple (the $4^+_1$ state continues this trend), the 
excited band does not show as much agreement; and given the above problem with the Hoyle state, one cannot be certain 
which, if either, calculation is better. For the $1^+$ states, the $1^+;1$ has good agreement between Cohen-Kurath and 
the NCSM, while the troublesome $1^+;0$ does not.  It will be particularly interesting to examine how the latter changes when an
\textit{ab initio} 3-body force is included in the calculation. 

\begin{figure}
\centering
\includegraphics[scale=0.45,clip]{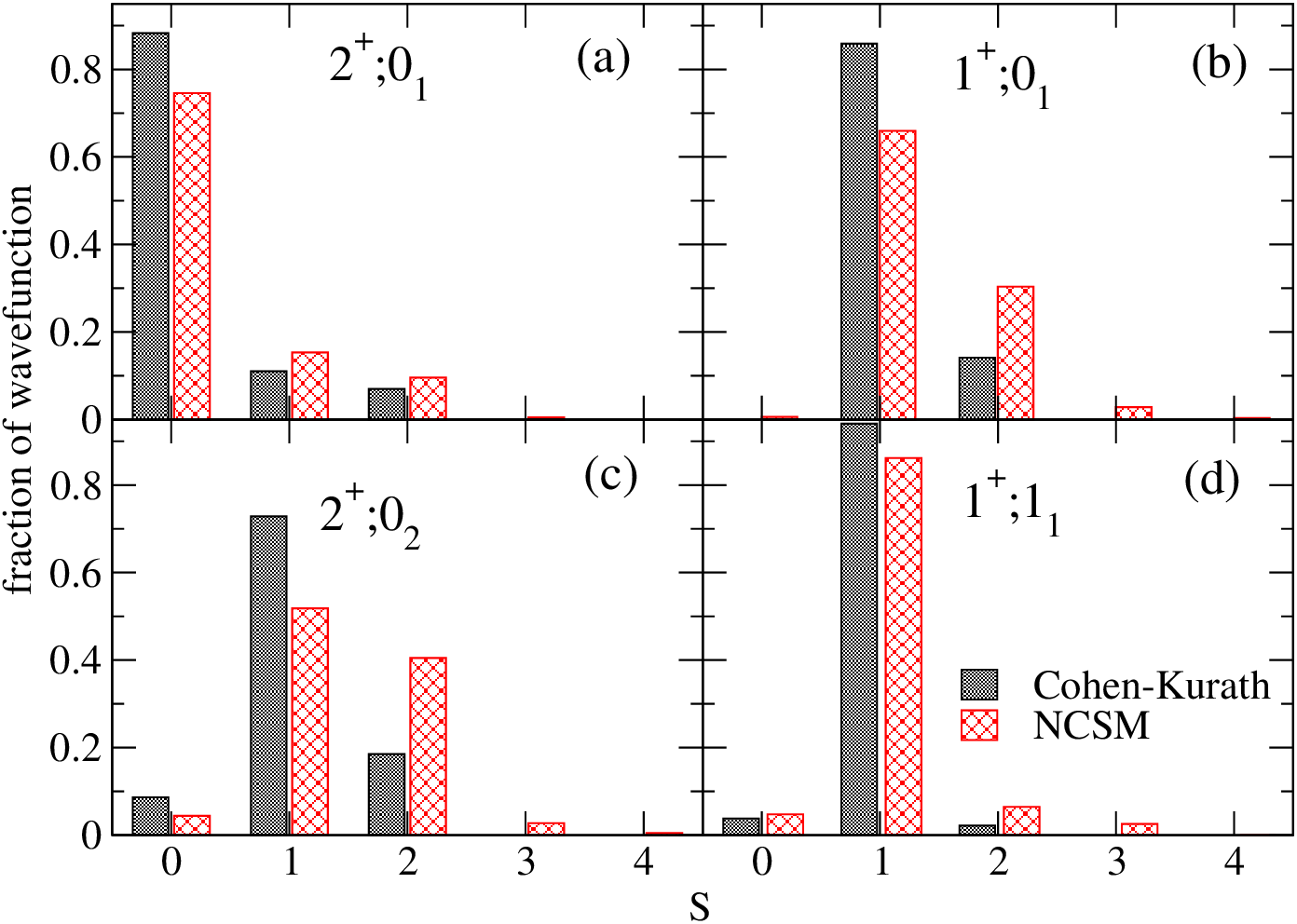}
\caption{(Color online) Decomposition of  low-lying states of $^{12}$C into 
components of good $S$ (total intrinsic spin, comparing wavfunction 
computed from
the Cohen-Kurath interaction (black/dark shaded), and from the NCSM (red/cross-hatched).}
\label{C12_Sdecomp}
\end{figure}

Given the importance and difficulty of this nuclide, I also show the $S$-decomposition in Fig.~\ref{C12_Sdecomp}.  The decomposition 
for the $0^+$ states must perforce mirror the $L$-decomposition, so I do not show those.  For the remaining states, the 
$S$-decomposition echoes that of the $L$-decomposition: the $2^+;0_1$ and $1^+;1$ show good agreement, while there are
significant discrepancies for the $2^+;0_2$ and $1^+;1$ states. 

Again, agreement does not mean the calculations are correct, nor does a discrepancy make any calculation illegitimate. 
After all, the $L$-$S$ decomposition is not something directly measurable by experiment. On the other hand, it is striking 
that the states with the clearest discrepancy between the two calculations are states known to be problematic in 
CI calculations, particular for the \textit{ab initio} NCSM.  Thus I suggest that $L$-$S$ decompositions can be useful in 
comparing and contrasting calculations. 

\subsection{$^9$Be and rotational bands}

\begin{figure}
\centering
\includegraphics[scale=0.45,clip]{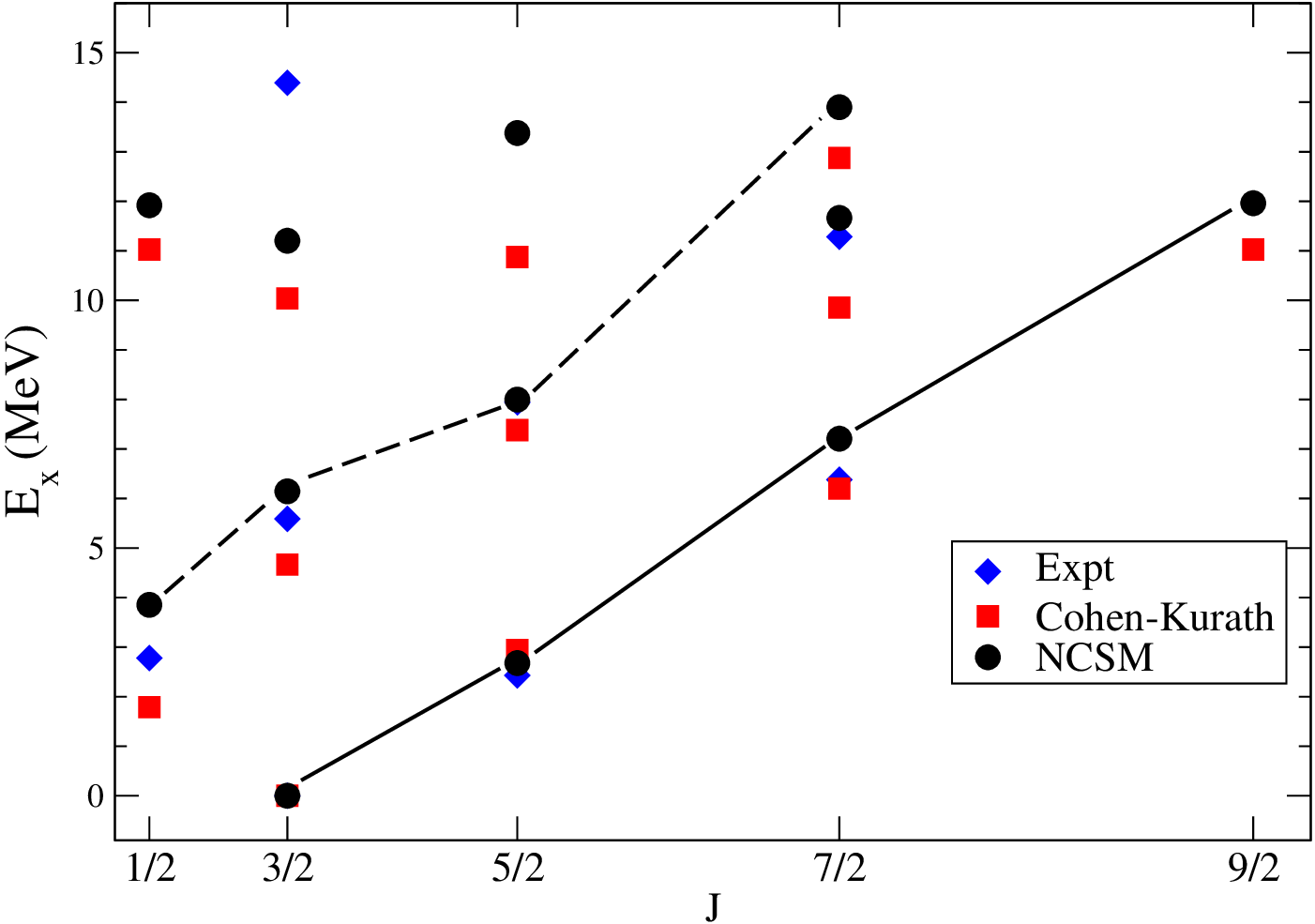}
\caption{(Color online) Low-lying excitation spectrum for $^9$Be, plotting excitation energy versus $J(J+1)$ to better 
illustrate rotational bands.  Shown are experimental data (blue diamonds) \cite{nndc}, phenomenological Cohen-Kurath 
calculations (red squares), and NCSM calculations (black circles); the latter were calculated with an SRG evolution parameter 
$\lambda = 2$ fm$^{-1}$ and a harmonic oscillator basis frequency of $\Omega = 20$ MeV. In order to guide the eye, I 
have added solid lines for the ground state NCSM band and dashed lines for the excited state NCSM band.}
\label{Be9rotational}
\end{figure}

\begin{figure}
\centering
\includegraphics[scale=0.45,clip]{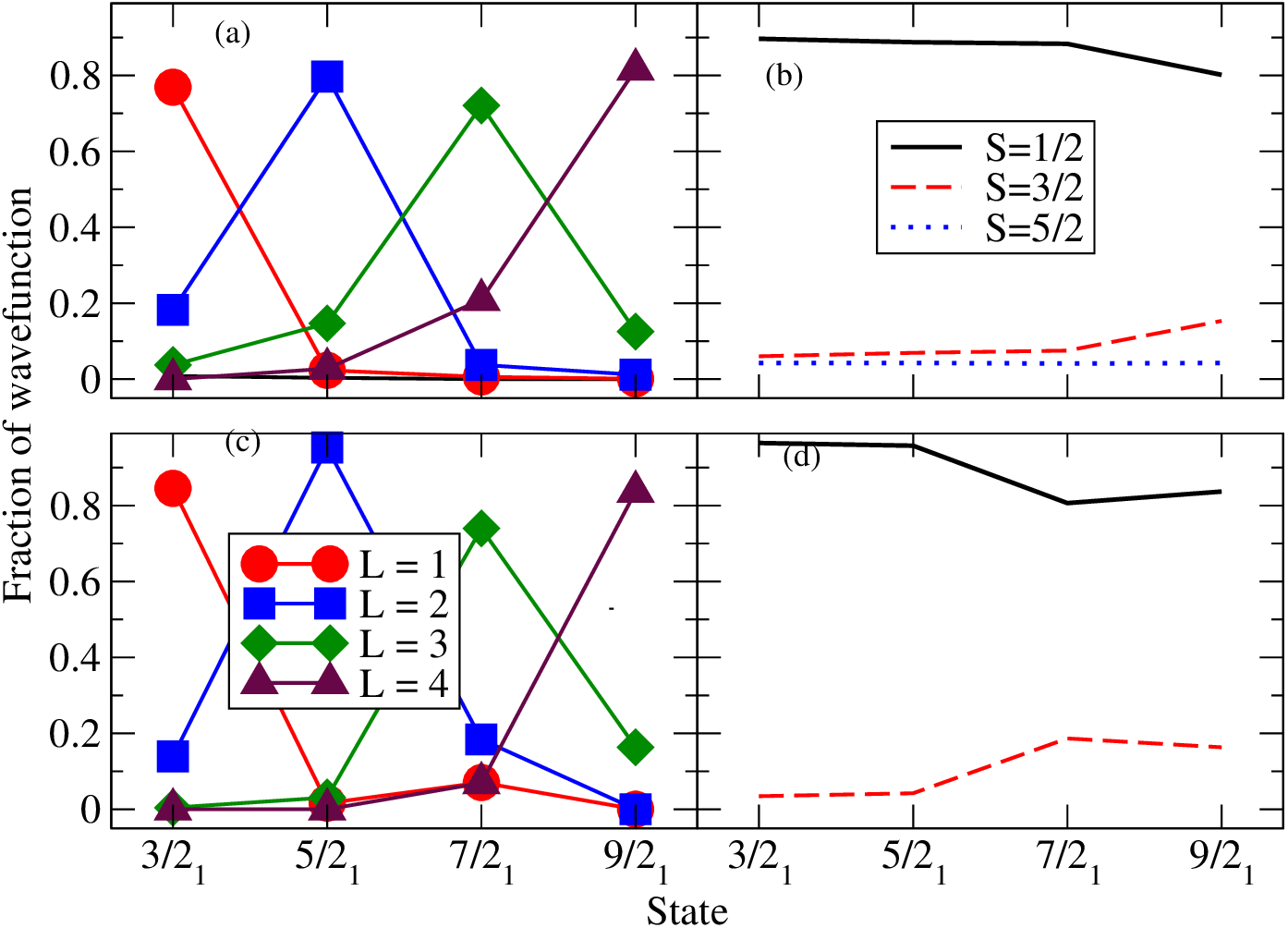}
\caption{(Color online) Decomposition of the ground-state rotation band for $^9$Be, for NCSM calculation (top row), 
with the parameters the same as in Fig.~\ref{Be9rotational}, and for Cohen-Kurath calculation (bottom row). 
 For the $L$-decomposition plots 
on the left, for each of the members 
of the rotational band $3/2^-_1$, $5/2^-_1$, $7/2^-_1$, and $9/2^-_1$, I give the fraction of the wavefunction with 
$L=1$ (red circles), $L=2$ (blue squares), $L=3$ (green diamonds), and $L=4$ (violet triangles).
For the $S$-decomposition plots on the right, I give the fraction of the wavefunction with $S=1/2$ (black solid line), $S=3/2$ 
(red dashed line), and $S=5/2$ (blue dotted line).
 }
\label{Be9gsband}
\end{figure}

\begin{figure}
\centering
\includegraphics[scale=0.45,clip]{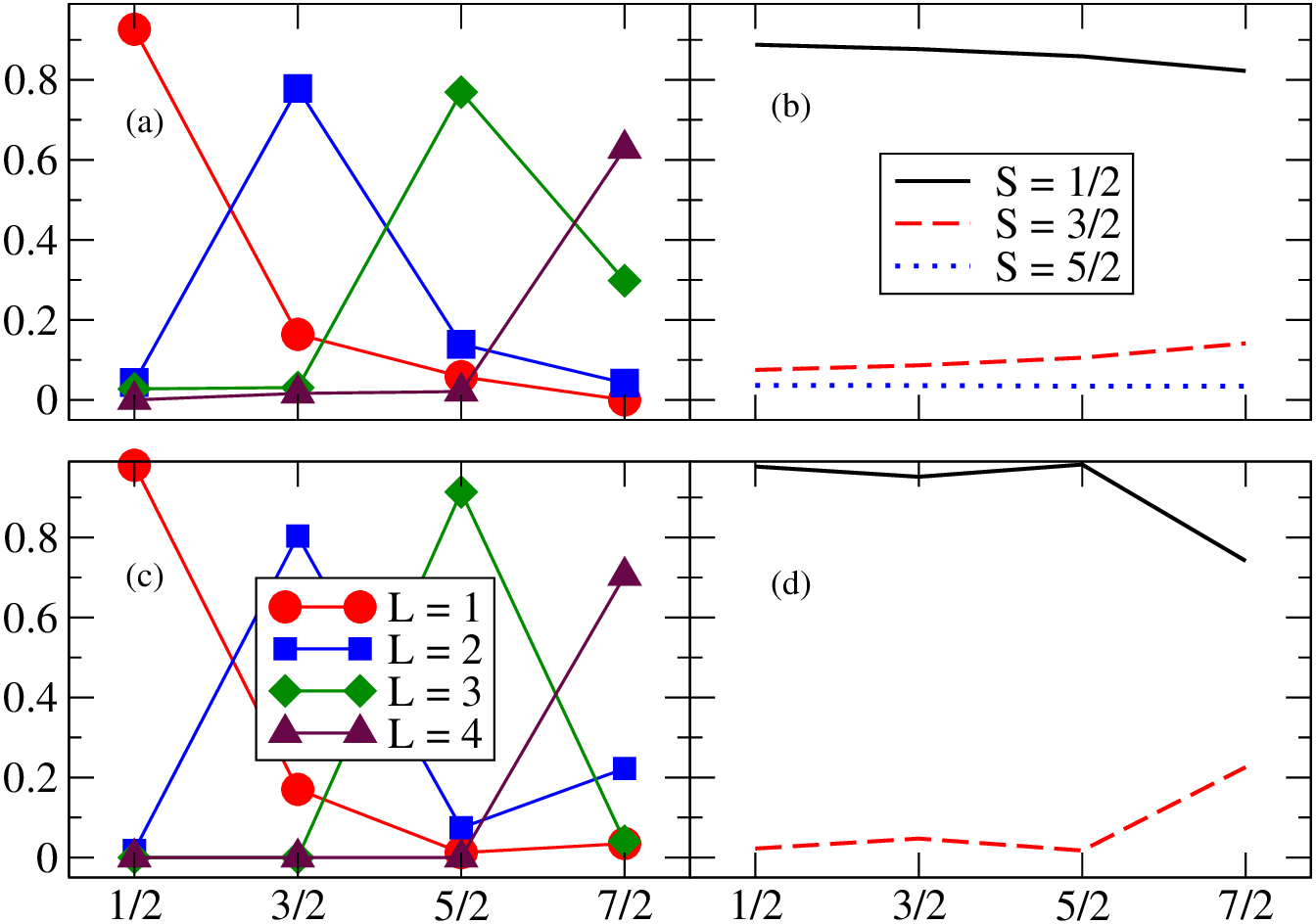}
\caption{(Color online) Decomposition of the excited-state rotation band for $^9$Be, for NCSM calculation (top row), 
with the parameters the same as in Fig.~\ref{Be9rotational}, and for Cohen-Kurath calculation (bottom row). 
 For the $L$-decomposition plots 
on the left, for each of the members 
of the rotational band $1/2^-_1$, $3/2^-_2$, $5/2^-_2$, and $7/2^-_3$, I give the fraction of the wavefunction with 
$L=1$ (red circles), $L=2$ (blue squares), $L=3$ (green diamonds), and $L=4$ (violet triangles).
For the $S$-decomposition plots on the right, I give the fraction of the wavefunction with $S=1/2$ (black solid line), $S=3/2$ 
(red dashed line), and $S=5/2$ (blue dotted line).}
\label{Be9exband}
\end{figure}

There have been recent studies of rotational band structure in NCSM calculations of light nuclides \cite{CMV13} using 
not only excitation spectra but also $E$2 and $M$1 transition strengths 
and electric quadrupole and magnetic dipole moments to identify band structure. 
As a complement to those studies, I use $L$-$S$ decomposition to analyze rotational bands.
Fig.~\ref{Be9rotational}  compares the low-lying excitation energies of $^9$Be
from experiment, from the Cohen-Kurath interaction in the $0p$ space, with a dimension of 62, and a NCSM calculation, 
with a dimension of 5.2 million, using $\lambda =2$ fm$^{-1}$ and $\hbar \Omega=20$ MeV. 
 All the low-lying states have $T=1/2$.
Following \cite{CMV13} the states are plotted with $J(J+1)$ along the $x$-axis and excitation energy along the $y$-axis, 
to better pick out rotational bands; I include lines for the reader's convenience.

Figs.~\ref{Be9gsband} and \ref{Be9exband} show the $L$- and $S$-decomposition for the ground-state and excited-state bands, 
respectively.  The top rows of plots are for the NCSM calculation, while the bottom rows are for the Cohen-Kurath calculations; 
one can see they are qualitatively indistinguishable.  The left-hand columns of plots are the $L$-decomposition. All these states have 
negligible $L=0$ fraction.  I show the fraction of the wavefunction with 
$L=1$ (red circles), $L=2$ (blue squares), $L=3$ (green diamonds), and $L=4$ (violet triangles). There is a clear evolution, as 
one would expect for a rotational band: in the ground state band,  $3/2^-_1$ dominated by $L=1$, the $5/2^-_1$ dominated by 
$L=2$,  $7/2^-_1$ by $L=3$, and $9/2^-_1$ by $L=4$, while for the excited state band $1/2^-_1$ is dominated by $L=1$, $3/2^-_2$ is dominated by 
$L=2$, $5/2^-_2$ dominated by 
$L=3$, and  $7/2^-_3$ by $L=4$. The right-hand columns of plots are the $S$-decomposition, showing the 
fraction of the wavefunction with $S=1/2$ (black solid line), $S=3/2$ 
(red dashed line), and $S=5/2$ (blue dotted line, only for the NCSM). The $S=1/2$ consistently dominates. These patterns 
are consistent with a particle-rotor picture. 

As with $^{11}$B, one can further decompose into proton and neutron contributions. In the rotational bands, the $S=1/2$ components 
are dominated ($> 95\%$) by $S_p = 0$, $S_n = 1/2$) while the $S=3/2$ components 
are dominated ($\geq 80\%$) by $S_p = 1$, $S_n = 1/2$). 

The reader will note  the excited-state band contains the $7/2^-_3$ state, not the $7/2^-_2$ state.  Caprio \textit{et al.} \cite{CMV13}
determined this on the basis of $B$(E2)s, $B(M1)$, and moments,  but here it becomes clear on the basis of the $L$-$S$ decomposition. 
Although I do not plot it, the $7/2^-_2$ state is dominated by $S=3/2$ rather than 1/2 for both NCSM and Cohen-Kurath, and by $L=2$ 
for the NCSM and by  a roughly equal mixture of $L=1$ and 2 for Cohen-Kurath wavefunctions rather than $L=4$ as 
found in the $7/2^-_3$ state, a clear violation of the rotational band pattern. 

Once again, the qualitative agreement between the NCSM and Cohen-Kurath calculations is striking. I propose $L$-$S$ decomposition 
as another tool for disentangling calculations of band structures. 

\subsection{Robustness}

Above I chose specific values of $\hbar\Omega$ for the harmonic oscillator basis, $N_\mathrm{max}$ for the 
truncation of the many-body basis, and  for the SRG evolution 
parameter $\lambda$.   These results are not very sensitive to the choice of these parameters, which 
can be demonstrated.

\begin{figure}
\centering
\includegraphics[scale=0.45,clip]{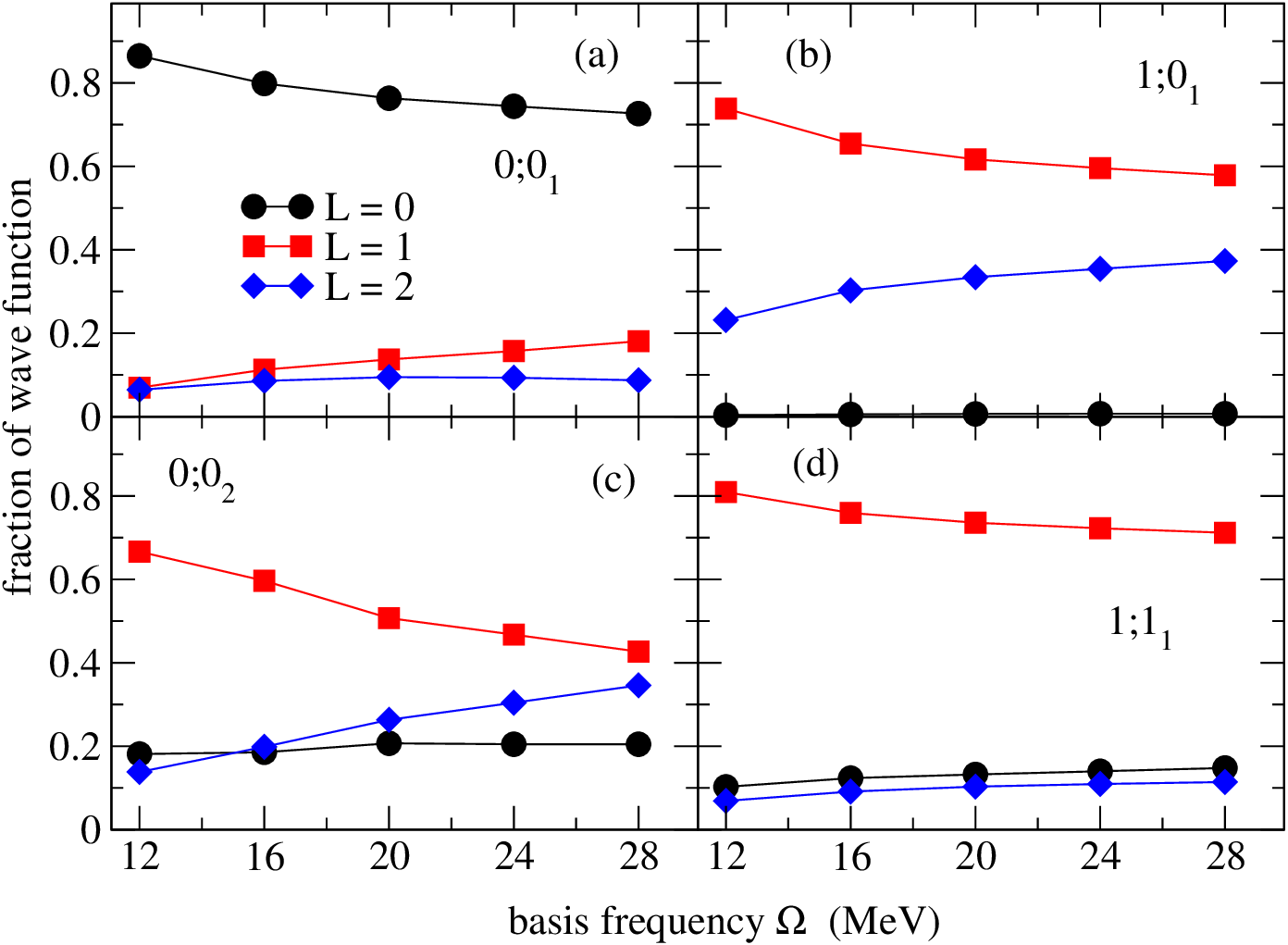}
\caption{(Color online) 
The $L$-decomposition for selected $^{12}$C NCSM states as a function of the harmonic oscillator basis frequency 
$\Omega$. The states are (a) $0;0_1$, (b) $1;0_1$; (c) $0;0_2$, and (d) $1;1_1$. Shown are the fraction of the wavefunctions 
for $L=0$ (black circles), $L=1$ (red squares), and $L=2$ (blue diamonds). 
}
\label{C12_LvsHW}
\end{figure}

  Starting with a baseline $^{12}$C with $N_\mathrm{max}=6$ and a baseline of $\lambda=2$ fm$^{-1}$ and 
$\hbar \Omega = 22$ MeV, 
 I first studied the dependence on the basis scaling. 
 Fig.~\ref{C12_LvsHW} shows how the $L$-decomposition
changes with  the basis frequency $ \Omega$ as it is varied from $ 12$ to $28$ MeV.
Although this corresponds to scaling the basis length parameter by a factor of 1.5, the 
decomposition is mostly robust,. 
Once again the states most sensitive are the problematic $1^+;0_1$ and the $0^+;0_2$ states;  
 in fact, with the latter the third $0^+;0$ state grows lower in energy as $\Omega$ increases and eventually switches places.  

\begin{figure}
\centering
\includegraphics[scale=0.45,clip]{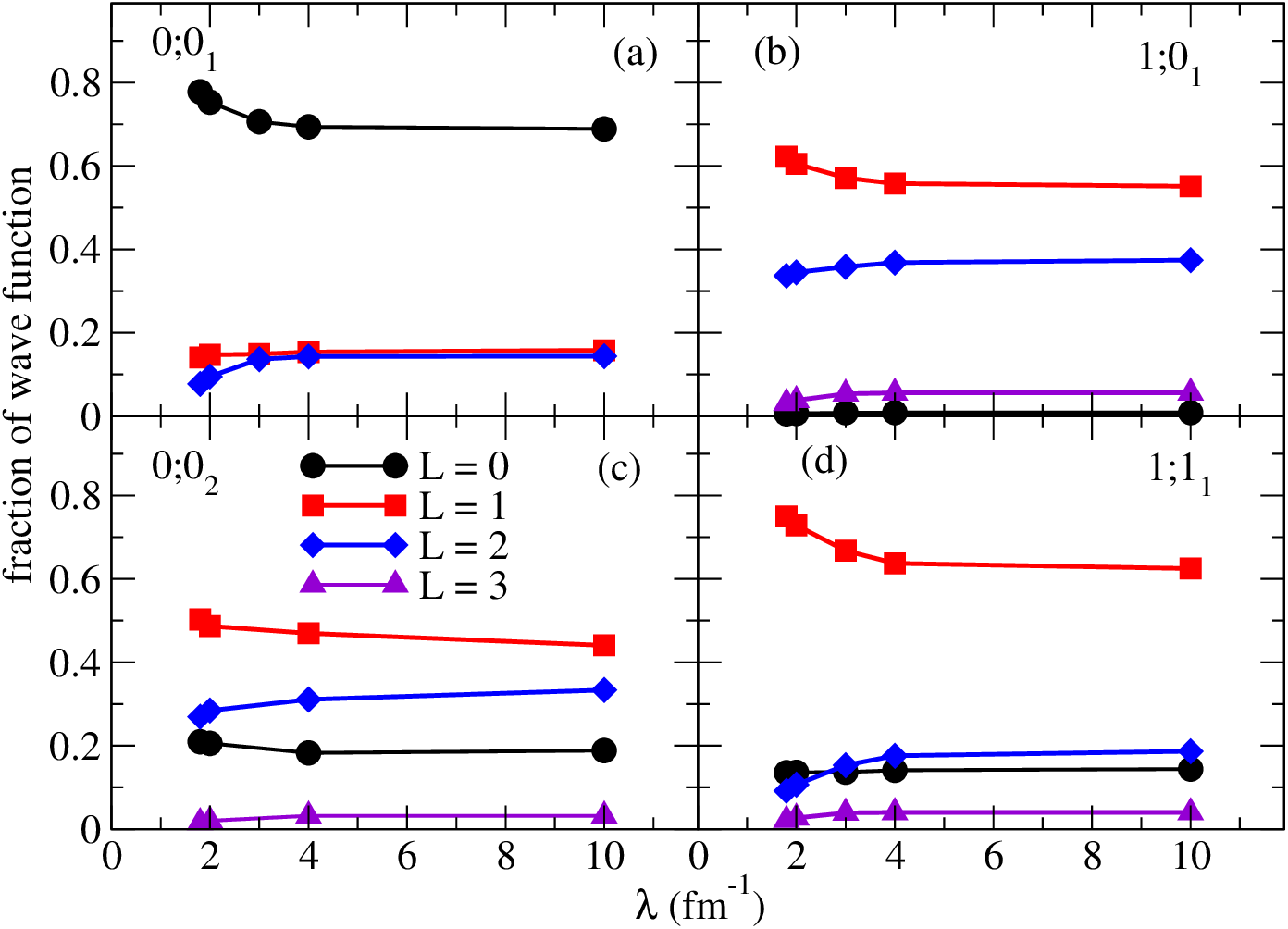}
\caption{(Color online) 
The $L$-decomposition for selected $^{12}$C NCSM states as a function of the SRG evolution parameter
$\lambda$. The states are (a) $0;0_1$, (b) $1;0_1$; (c) $0;0_2$, and (d) $1;1_1$. Shown are the fraction of the wavefunctions 
for  $L=0$, (black circles), $L=1$ (red squares),  $L=2$ (blue diamonds) and $L=3$ (violet triangles). 
}
\label{C12_LvsLam}
\end{figure}

The second study, Fig.~\ref{C12_LvsLam} was the dependence of the $L$-decomposition on the SRG evolution parameter $\lambda$, as it goes from 
10 fm$^{-1}$, which is almost the bare interaction, down to 1.8 fm$^{-1}$. Values in the range 1.8 to 2.2 fm$^{-1}$ are 
typically used for NCSM calculations. Although there is some evolution  as $\lambda$ goes below 4  fm$^{-1}$, overall 
the dependence on $\lambda$ is  modest. 

\begin{figure}
\centering
\includegraphics[scale=0.45,clip]{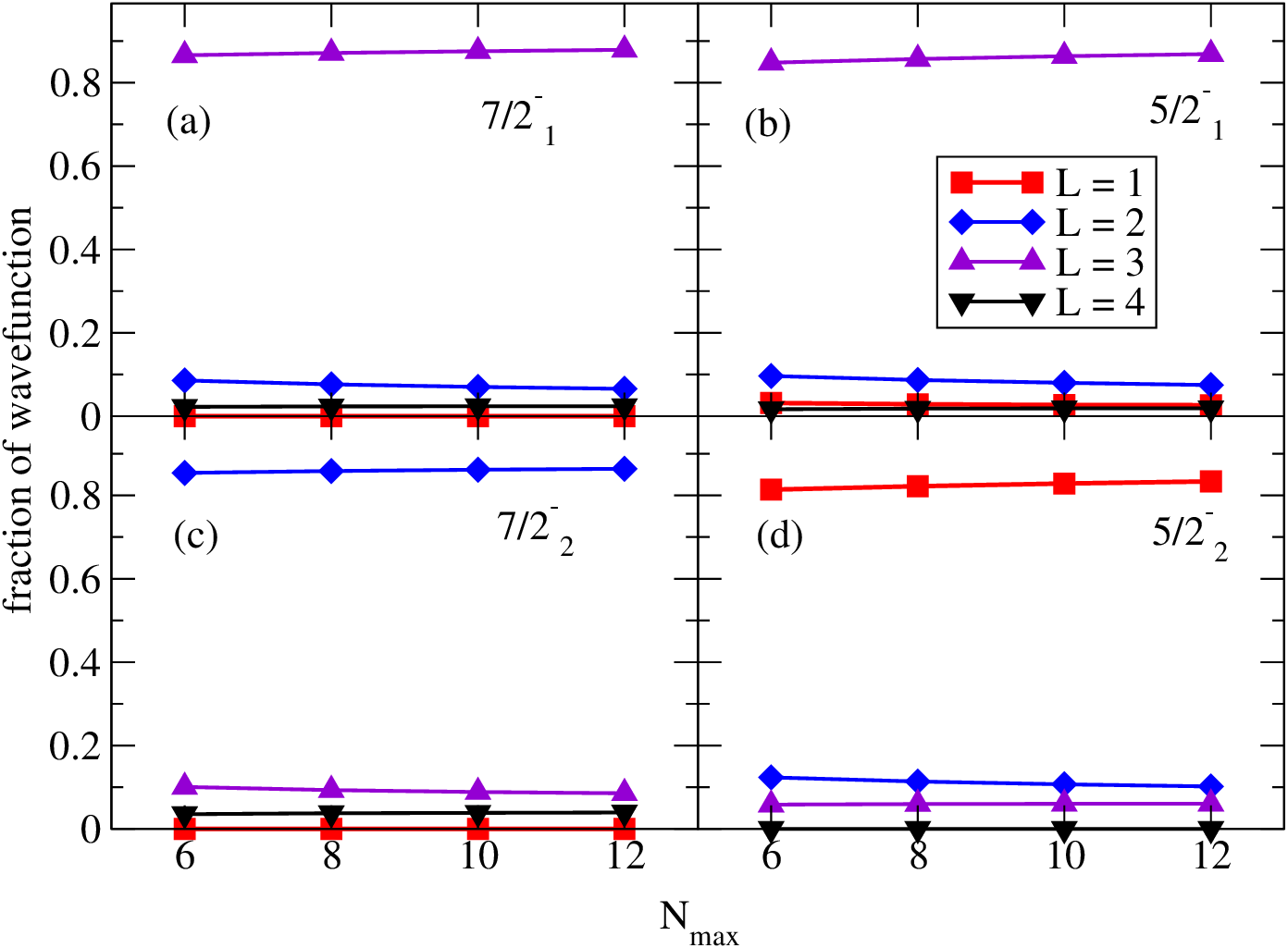}
\caption{(Color online) 
The $L$-decomposition for selected $^{7}$Li NCSM states as a function of model space truncation $N_\mathrm{max}$, with 
SRG evolution parameter
$\lambda$ fixed at 2 fm$^{-1}$ and the harmonic oscillator basis frequency fixed at 22 MeV. All states have $T=1/2$. 
 The states are (a) $7/2^-_1$, (b) $5/2^-_1$; (c) $7/2^-_2$, and (d) $5/2^-_2$. Shown are the fraction of the wavefunctions 
for $L=1$ (red squares),  $L=2$ (blue diamonds),  $L=3$ (violet up triangles), and $L=4$ (black down triangles).
}
\label{Li7LvNmax}
\end{figure}

Finally, I studied how well the $L$-decomposition had converged in $N_\mathrm{max}$.  Above, in section \ref{b10} I already demonstrated the the 
$L$-decomposition is unchanged for $^{10}$B as one goes from $N_\mathrm{max}=6$ to 8, even though the ground state angular momentum changes. 
 To study a broader range of $N_\mathrm{max}$ , I chose a lighter system,
$^7$Li, where I could compute models spaces up from $N_\mathrm{max}$ (dimension =663,527) up to  $N_\mathrm{max}=12$ 
(dimension =252 million)
on a desktop computer. 
 Fig.~\ref{Li7LvNmax} shows the $L$-decomposition does not change much even as the model 
space increases nearly three orders of magnitude. 

As a final note, the mirror nuclide $^7$Be has been identified as having a  rotational band in the yrast $1/2_1$, $3/2_1$, $5/2_1$, $7/2_1 \ldots$ states \cite{CMV13}.  
I find these 
states   dominated by 
$S=1/2$, though the $1/2_1$ and $3/2_1$ states are dominated by $L=1$ while the $5/2_1$, $7/2_1$ states dominated by $L=3$, as 
seen above. This differs from the rotational structure seen in $^9$Be above, where the states in both the ground and excited bands (Figs.~\ref{Be9gsband},\ref{Be9exband}) 
are dominated by $L=1,2,3,4$, successively.  This may be due to the difference between the chiral nucleon-nucleon force used here \cite{EM03} and the JISP16 force 
\cite{JISP16} used in \cite{CMV13} (although I include isospin breaking in my calculation, both $^7$Li and $^7$Be yield very similar results); but also, of all the Be isotopes in \cite{CMV13}, the ground band of $^7$Be exhibits the most 
irregular behavior with regards to the magnetic dipole moment and M1 transition strengths. ( I find the $3/2_2$, $5/2_2$, and $7/2_2$ are all  dominated by $S=3/2$, with 
$L= 1$ for $3/2_2$, $5/2_2$, and $L=2$ for $7/2_2$; a second band is not investigated in \cite{CMV13}.)
Further investigation may be warranted in future. 



\section{Conclusions and acknowledgements} 

I have taken NCSM wave functions computed with \textit{ab initio} two-body interactions from chiral effective field theory, 
softened with SRG, and decomposed them into their $L$ (total orbital angular momentum) and
$S$ (total spin) components, for selected $p$-shell nuclides.  
 Somewhat remarkably, there is rather good agreement 
with the decomposition of wavefunctions for the same nuclides using the phenomenological Cohen-Kurath force, despite 
vastly different origins and nearly six orders of magnitude difference in the dimensions of the model spaces.  I think this 
helps assure us, if we need such assurance, that both the old guard and the new have mostly captured the correct physics. 

As examples of the utility of $L$-$S$ decomposition, I looked at states in $^{12}$C known to be difficult to calculate, where 
the strongest discrepancies between the NCSM and Cohen-Kurath wavefunctions showed up, and also showed how in $^9$Be rotational band structure showed with a clear pattern consistent with 
a particle-rotor model. It will be interesting in the future to investigate other rotational bands in more detail, in particular $^7$Li/Be. 

The results are fairly robust even when the basis is changed, and not very sensitive to the SRG evolution--although 
there is some sensitive at the end of SRG evolution.  It will be very interesting therefore to use three-body forces, both 
``true'' three-body forces and those induced by SRG. The former are known to affect spin-orbit coupling, and it will be interesting 
to see if it brings the NCSM results closer to or further apart from the Cohen-Kurath; the latter should decrease sensitivity 
to SRG evolution.  In principal of course, one should also evolve the $\hat{L}^2$ and $\hat{S}^2$ operators which 
should also decrease sensitivity to SRG evolution \cite{SQJ14}.  

Therefore an important future step will be to look at chiral interactions including \textit{ab initio} three-body forces 
\cite{JNF09,JMF13}, and 
alternate \textit{ab initio} approaches such as the JISP16 interaction\cite{JISP16} which has off-shell matrix elements 
tuned to best match binding energies, and thus reduce the need for three-body forces.

I thank P. Navr\'atil and R. Roth for helpful and encouraging discussions, and P. Navr\'atil for the code generating and 
evolving via SRG the N3LO nucleon-nucleon matrix elements.

This material is based upon work supported by the U.S. Department of Energy, Office of Science, Office of Nuclear Physics, 
under Award Number  DE-FG02-96ER40985.  Additional supercomputing support for this work came from the Lawrence LIvermore National Laboratory 
institutional Computing Grand Challenge program.


\begin{thebibliography}{99}

\bibitem{May48} M. Mayer, Phys. Rev. \textbf{74}, 235 (1948); Phys. Rev. \textbf{75}, 1969 (1949).

\bibitem{HJS49} O. Haxel, J. H. D. Jensen, and H. E. Suess Phys. Rev. \textbf{75}, 1766 (1949)

\bibitem{FHN49} E. Feenberg, K. C. Hammack, and L. W. Nordheim, Phys. Rev. \textbf{75}, 1968 (1949).

\bibitem{CarlsonGFMC} J. Carlson,  Phys. Rev. C \textbf{36}, 2026 (1987); Nucl. Phys. \textbf{A 508}, 141 (1990).
%
\bibitem{GFMC} S. C. Pieper and R. B. Wiringa, Annu. Rev. Nucl. Part. Sci \textbf{51}, 53 (2001).
%
\bibitem{DHJ04} D. J. Dean and M. Hjorth-Jensen, Phys. Rev. C \textbf{69}, 054320 (2004).
%
\bibitem{HPDHJ10} G. Hagen, T. Papenbrock, D. J. Dean, and M. Hjorth-Jensen, Phys. Rev. C \textbf{82}, 034330 (2010).
\bibitem{NVB00} P. Navr\'atil, J. P. Vary, and B. R. Barrett, Phys. Rev. C \textbf{62}, 054311 (2000).
\bibitem{Edmonds} A. R Edmonds, Angular momentum in quantum mechanics (Princeton University Press, Princeton, 1960).

\bibitem{FW37}  E. Feenberg and E. Wigner, Phys. Rev. \textbf{51}, 95 (1937).

\bibitem{FP37}  E. Feenberg and M. Phillips, Phys. Rev. \textbf{51},597 (1937).

\bibitem{Flo52} B. H. Flowers,  Proceedings of the Royal Society of London, A \textbf{212}, 248 (1952).

\bibitem{Kur52} D. Kurath, Phys. Rev. \textbf{88}, 804 (1952).
\bibitem{BG77} P.J. Brussard and P.W.M. Glaudemans, \textit{Shell-model applications 
in nuclear spectroscopy} (North-Holland Publishing Company, Amsterdam 1977).
%
\bibitem{BW88}  B. A. Brown and B. H. Wildenthal, Annu. Rev. Nucl. Part. Sci. 38, 29 (1988).
%
\bibitem{ca05}  E. Caurier, G. Mart\'{\i}nez-Pinedo, F. Nowacki, A. Poves, and A. P. Zuker, Rev. Mod. Phys.
\textbf{77}, 427 (2005).
\bibitem{Pov01} A. Poves, J. Sanchez-Solano, E. Caurier, and F. Nowacki, Nucl. Phys. \textbf{A694}, 157 (2001).
%
\bibitem{KB68} T. T. S. Kuo and G. E. Brown, Nucl. Phys. \textbf{A114}, 241 (1968).
%
%
\bibitem{USDB} B. A. Brown and W. A. Richter, Phys. Rev. C \textbf{74}, 034315 (2006). 
%
\bibitem{CK65}  S. Cohen and D. Kurath, Nucl. Phys. \textbf{73}, 1 (1965).






%
\bibitem{Kur56} D. Kurath, Phys. Rev. \textbf{101}, 216 (1956).
%
\bibitem{Ing53} D. R. Inglis, Rev. Mod. Phys. \textbf{25}, 390 (1953).
\bibitem{Mil09} D. J. Millener, in XVIIIth Indian Summer School in Physics, Topics in 
strangeness nuclear physics, P. Bydzovsky, A. Gal, and J. Mares, eds.,
 Lect. Notes Phys. \textbf{724}, 31 (2007), also arxiv:0902.2142.

\bibitem{NO03} P. Navr\'atil and W. E. Ormand, Phys. Rev. C \textbf{68}, 034305 (2003).
%
\bibitem{BIGSTICK} C. W. Johnson. W. E. Ormand, and P. G. Krastev, Comp. Phys. Comm. \textbf{184},
2761 (2013).
\bibitem{GVL} G. H. Golub and C. F. van Loan, \textit{Matrix computations}, 3rd ed.
(The Johns Hopkins University Press, Baltimore, 1996).
%
\bibitem{Lanczos} R. R. Whitehead, A. Watt, B. J. Cole, and I. Morrison,  Adv. Nucl. Phys. \textbf{9}, 123 (1977).
%
\bibitem{Pal67} F. Palumbo, Nucl. Phys. \textbf{A 99}, 100 (1967).
%
\bibitem{PP68} F. Palumbo and D. Prosperi, Nucl. Phys. \textbf{A 115}, 296 (1968).
%
\bibitem{GL74} D. H. Gloeckner and R. D. Lawson, Phys. Lett. \textbf{B 53}, 313 (1974).
\bibitem{Argonne} R. B. Wiringa, V. G. J. Stoks, and R. Schiavilla, Phys. Rev. C \textbf{51}, 38 (1995). 
%
\bibitem{EM03} D. R. Entem and R. Machleidt, Phys. Rev. C \textbf{68}, 041001 (2003).
%
%
\bibitem{Wei90} S. Weinberg, Phys. Lett. \textbf{B 251}, 288 (1990).
%
\bibitem{Wei91} S. Weinberg, Nucl. Phys. \textbf{B 363}, 3 (1991).
%
\bibitem{OvK92} C. Ord\'o\~nez and U. van Kolck, Phys. Lett. \textbf{ B 291}, 459 (1992).
%
\bibitem{BvK02} P. F. Bedaque and U. van Kolck, Annu. Rev. Nucl. Part. Sci. \textbf{52}, 339 (2002).
%
%
\bibitem{Sh98} I. Shavitt, Mol. Phys. \textbf{94}, 3 (1998).
%
%
\bibitem{GW93} S. D. Glazek and K. G. Wilson, Phys. Rev. D \textbf{48}, 5863 (1993).
%
\bibitem{Weg94} F. Wegner, Ann. Phys. \textbf{506}, 77 (1994).
%
\bibitem{BFP07} S. K. Bogner, R. J. Furnstahl, and R. J. Perry, Phys. Rev. C. \textbf{75} 061001 (2007).
%
\bibitem{BFS10} S. K. Bogner, R. J. Furnstahl, and A. Schwenk, Prog. Part. Nucl. Phys. \textbf{65}, 94 (2010).
%
\bibitem{JNF09}  E. D. Jurgenson, P. Navr\'atil, and R. J. Furnstahl, Phys. Rev. Lett. \textbf{103}, 082501 (2009).

\bibitem{JMF13} E. D. Jurgenson, P. Maris, R. J. Furnstahl,  P. Navr\'atil \textit{et al.}, Phys. Rev. C \textbf{87}, 054312 (2013).
%
\bibitem{CPZ90} E. Caurier, A. Poves, and A. P. Zuker, Phys. Lett. \textbf{B252},  13 (1990);
Phys. Rev. Lett.  \textbf{74}, 1517 (1995).
%
\bibitem{CMN99} E. Caurier, G. Mart\'inez-Pinedo, F. Nowacki, A. Poves, \textbf{et al.}, Phys. Rev. C \textbf{59}, 2033 (1999).
%
\bibitem{HNZ05} W. C. Haxton, K. M. Nollett, and K. M. Zurek, 
 Phys. Rev. C 72, 065501 (2005).
%
\bibitem{GDJ00} V. G. Gueorguiev,  J. P. Draayer, and C. W. Johnson, Phys. Rev. C \textbf {63}, 014318 (2000).
%
\bibitem{nndc}
{\tt http://www.nndc.bnl.gov}
%
\bibitem{NGV07} P. Navr\'atil, V. G. Gueorguiev, J. P. Vary, W. E. Ormand, and A. Nogga, Phys. Rev. Lett. \textbf{99}, 042501 (2007).

%
\bibitem{Hoyle} F. Hoyle, Astrophys. J. Suppl. Ser. \textbf{1}, 121 (1954).
%
\bibitem{Neff12} T. Neff, Proceedings of Horizons of Innovative Theories, Experiments, and 
Supercomputing in Nuclear Physics, K. D. Launey, M. A. Caprio, J. E. Escher, J. G. Hirsch, and C. W. Johnson, eds., Journal Of Physics:Conference Series \textbf{403}, 012028 (2012).
\bibitem{MVC14} P. Maris, J. P. Vary, A. Calci, J. Langhammer, \textit{et al.}, Phys. Rev. C \textbf{90}, 014314 (2014).
\bibitem{CMV13} M. A. Caprio, P. Maris, and J. P. Vary, Phys. Lett. \textbf{B 719}, 179 (2013);  P. Maris, M. A. Caprio, and J. P. 
Vary, Phys. Rev. C \textbf{91}, 014310 (2015) .

\bibitem{JISP16}  A. M. Shirokov, J. P. Vary, A. I. Mazur, and T. A. Weber, Phys. Lett. \textbf{B644}, 33 (2007). 

\bibitem{SQJ14} M. D. Schuster,  S. Quaglioni, C. W. Johnson, E. D. Jurgenson, and P. Navr\'atil, Phys. Rev. C \textbf{90}, 
 011301(R) (2014).


\end{thebibliography}
\end{document}